\documentclass[prl,twocolumn,showpacs,amsmath,amssymb,superscriptaddress]{revtex4}
\usepackage{amssymb}
\usepackage{amsmath}
\usepackage{colordvi}
\usepackage[colorlinks]{hyperref}
\usepackage{amsthm}
\usepackage{subeqnarray}
\usepackage{cases}
\usepackage{enumerate}
\usepackage{graphicx}
\usepackage{epsfig}
\usepackage{epstopdf}
\usepackage{subfigure}
\usepackage{color}
\usepackage{CJKutf8}

\def\avg(#1){\langle#1\rangle}

\def\be{\begin{equation}}
\def\ee{\end{equation}}
\def\bea{\begin{eqnarray}}
\def\eea{\end{eqnarray}}

\begin{document}
\title{Stable diagonal stripes in the t-J model at $\bar{n}_h$=1/8 doping from fPEPS calculations}

\author{Shao-Jun Dong}
\affiliation{CAS Key Laboratory of Quantum Information, University of Science and Technology of China, Hefei 230026, People's Republic of China}
\affiliation{Synergetic Innovation Center of Quantum Information and Quantum Physics, University of Science and Technology of China, Hefei 230026, China}

\author{Chao Wang}
\affiliation{CAS Key Laboratory of Quantum Information, University of Science and Technology of China, Hefei 230026, People's Republic of China}
\affiliation{Synergetic Innovation Center of Quantum Information and Quantum Physics, University of Science and Technology of China, Hefei 230026, China}

\author{Yong-Jian Han}
\email{smhan@ustc.edu.cn}
\affiliation{CAS Key Laboratory of Quantum Information, University of Science and Technology of China, Hefei 230026, People's Republic of China}
\affiliation{Synergetic Innovation Center of Quantum Information and Quantum Physics, University of Science and Technology of China, Hefei 230026, China}

\author{Chao Yang}
\affiliation{Computational Research Division, Lawrence Berkeley National Laboratory, Berkeley, CA 94720, USA}

\author{Lixin He}
\email{helx@ustc.edu.cn}
\affiliation{CAS Key Laboratory of Quantum Information, University of Science and Technology of China, Hefei 230026, People's Republic of China}
\affiliation{Synergetic Innovation Center of Quantum Information and Quantum Physics, University of Science and Technology of China, Hefei 230026, China}

\begin{abstract}
We investigate the 2D t-J model  at a hole doping of $\bar{n}_h$=1/8 using recently developed high accuracy
fermionic projected entangled pair states(fPEPS) method.
By applying stochastic gradient descent method combined with Monte Carlo sampling technique,
we obtain the ground state hole energy $E_{\rm hole}$=-1.6186 for $J/t$=0.4.
We show that the ground state has stable diagonal stripes instead of vertical stripes with width of 4 unit cells,
and stripe filling $\rho_l$=0.5.
We further show that the long range superconductivity order is suppressed at this point.
\end{abstract}
\maketitle


The high-Tc superconductivity~\cite{Bednorz1986,Wu1987} is probably one of the most exciting and also challenging open problems in condensed matter physics.
The strong coupling between the spin and charge degrees of freedom leads
to various competing orders at low temperature, resulting in rich phase diagrams\cite{Damascelli2003}.
Specifically, the hole doping near $\bar{n}_h$=1/8 provides an ideal system to experimentally study the ground state of the pseudogap \cite{Valla2006} which is one of the most salient phenomena in high-Tc superconductivity\cite{Lee2006}.
Near this doping level, the charge and spin stripe orders are observed in some cuprate compounds, e.g., La$_{1.875}$Ba$_{0.125}$CuO$_4$, by various experimental techniques, including angle-resolved photoemission and scanning tunneling microscopy \cite{Valla2006}, neutron and x-ray scattering \cite{Ichikawa2000},
etc.

It is widely believed that the physics of superconductivity could be understood as doped Mott insulators\cite{Lee2006}, which could be described by the 2-dimensional Hubbard model~\cite{Hubbard1963,ANDERSON1987} and the
$t$-$J$ model~\cite{Zhang1988}, the strong coupling limit of Hubbard model.
%
However, the theoretical results about the ground state near hole doping $\bar{n}_h$=1/8 in the $t$-$J$ model are still highly controversial \cite{Han2001,Sherman2003,Capello2008}. The question about whether the ground state has the stable stripe order, and the relation between the superconductivity and the stripe order are under intensive debates.
\cite{Hellberg1999,White1998,Tranquada1997,Hu2012,Corboz2011,Corboz2014,Han2001,Capello2008,Sherman2003}
Early works on this issue have been reviewed in Refs.\onlinecite{Kivelson2003,Carlson2003}.
 Very recently, variational quantum Monte Carlo (vQMC) simulations combined few Lanczos steps \cite{Hu2012}, suggest that the ground state at $\bar{n}_h$=1/8 is homogeneous without stripes order. These results are contradictory to the results of the early DMRG calculations \cite{White1998,White2000PSAndStripe,White2004,White2009ttJ}. More recently, iPEPS with full update calculations ~\cite{Corboz2011,Corboz2014}  suggest that the ground state has stable stripes. Nevertheless,  the calculations ~\cite{Corboz2011,Corboz2014} suggest that the uniform phase is energetically very close to the stripe phase, and the energy difference become even smaller with increasing bond dimension. Therefore, it is hard to determine what the true ground state is unless fully converged calculations are performed.

The projected entangled pair states method (PEPS),\cite{Schollwoeck2011,Verstraete2008,Xiang2008,Verstraete2004,Cirac2010,Verstraete06} and its generalization to
fermionic systems (fPEPS) \cite{Gu2010,Gu2013,Corboz2010,Kraus2010} provide systematically improvable variational wave functions for many-body problems.
In recent works, we developed a gradient method combined with Monte Carlo sampling techniques to optimize the (f)PEPS wave functions with controlled accuracy\cite{Liu2017,He2018,Dong2019}. This method significantly reduces the scaling with respect to the bond dimension $D$, thereby allowing a much larger bond dimension to be used, resulting in highly accurate and converged results for large finite systems.
In this work, we apply this recently developed and highly accurate fPEPS method to explore the true ground state of the $t$-$J$ model at doping level $\bar{n}_h\sim 1/8$. The computational results allow us to shed some new light on this long standing open problem.
From our computation we obtained the hole energy $E_{\rm h}$= -1.6186 for $J/t$=0.4 in the thermodynamic limit.
Remarkably, we find that the ground state of the $t$-$J$ model at $1/8$ hole doping has stable stripes that are along the diagonal directions instead of the vertical direction suggested by previous works\cite{White1998,Corboz2011,Corboz2014,Han2001,Normand2001,Capello2008}
with stripe hole filling $\rho_l$=0.5.
We further show that the long range superconductivity order at this point is suppressed.


The $t$-$J$ model is defined on a two-dimensional square lattice as,
\begin{equation}
H=-t\sum_{\langle i,j \rangle,\sigma}(c_{i,\sigma}^\dag c_{j,\sigma} + H.c.) + J\sum_{\langle i,j \rangle}({\bf S}_i \cdot {\bf S}_j-\frac{1}{4}n_i n_j)
\end{equation}
where $\langle i,j\rangle$ are the nearest-neighbor sites.  $c_{i,\sigma}$ ($c_{i,\sigma}^\dag$) is the electron annihilation (creation) operator of spin $\sigma$ ($\sigma=\uparrow,\downarrow$) on site $i$,
whereas $n_i=\sum_\sigma c_{i,\sigma}^\dag c_{i,\sigma}$ and $\vec{S}_i$ are the electron number and the spin-1/2 operators respectively. Double occupations are not allowed.

We solve the model by using recently developed fPEPS~\cite{Barthel2009,Kraus2010,Gu2010,Corboz2010,Corboz2011,Gu2013,Dong2019}
 method.
The fPEPS wave functions are first optimized via an imaginary time evolution with simple update (SU) \cite{Xiang2008}
scheme, followed by gradient optimization combined with Monte Carlo sampling techniques.\cite{Liu2017,Dong2019}
U(1) symmetry is enforced during the calculations to conserve the number of electrons in the system.
More details of the methods are discussed in Refs.~\onlinecite{Liu2017,He2018,Dong2019}.
It is well known that the environment effects are oversimplified
in the SU method. Therefore, the use of SU may introduce large errors. However, the results can be used as a good starting point for the subsequent gradient optimization.
The gradient optimization method treats the environment effects exactly with controllable errors, and therefore can obtain much more accurate results.
The Monte Carlo sampling techniques \cite{Foulkes2001,Sandvik2007} are used to calculate the energies and their gradients, which may greatly reduce the complexity of the calculations, and allow us to use large virtual bond dimension $D$ and truncation parameters $D_c$ to converge the results.\cite{Dong2019}
In this work, open boundary conditions (BC) are used. We focus on the parameters of $t$=1 and $J/t=0.4$ with hole doping of $\bar{n}_h=\frac{1}{8}$.

%

\begin{figure} [tbp]
		\begin{center}
\includegraphics[width=0.45\textwidth]{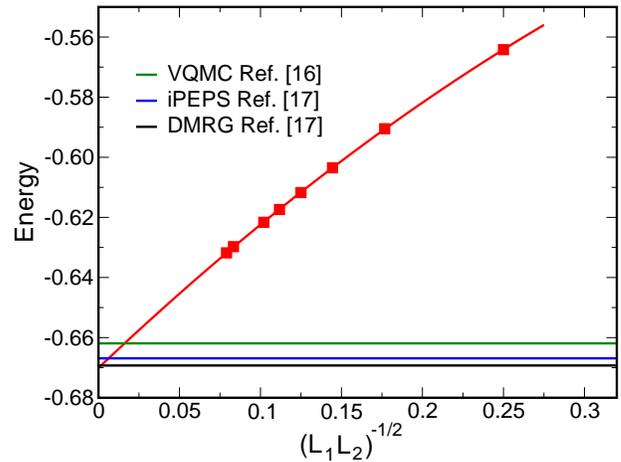}
		\caption{The ground state energies of t-J model with $t$=1, and $J/t=0.4$ at hole doping $\bar{n}_h=\frac{1}{8}$. The red squares represent the energies on different lattice sizes. The energies are extrapolated to the thermodynamic limit via a second-order polynomial function of $\sqrt{L_1L_2}$.
The green, blue and black lines are the ground state energies obtained by QMC, iPEPS simple update and DMRG methods respectively.}\label{pic:Scaling}
		\end{center}
\end{figure}

Figure~\ref{pic:Scaling} depicts the ground state energies of $t$-$J$ model at hole doping $\bar{n}_h$=1/8,
for different system sizes $L_1\times L_2$, ranging from 4$\times$4 to 12$\times$12.
In all calculations, the virtual bond dimension $D$ is fixed to 12, whereas the truncated dimension is set to $D_c$=48 to ensure that the energies are well converged at the given $D$.
As shown in our previous work,\cite{Dong2019} for t-J model at $\bar{n}_h$=1/8, usually the choice of $D_c$=3$D$=36 can ensure convergence, which is only weakly size dependent.
For the 4$\times$4 system, the ground state energy obtained by our calculation
is $E_{\rm fPEPS}$=-0.56428, compared with the exact energy $E_{\rm ED}$=-0.56436,
where the energy difference is about 1$\times 10^{-4}$.

We extrapolate the ground state energies to the thermodynamic limit using a second-order
polynomial fitting on $\sqrt{L_1L_2}$.
The extrapolated ground state energy in the thermodynamic limit is $E_\infty=$-0.6701.
The corresponding energy per hole is defined as $E_{\rm hole}=[E(\bar{n}_h)-E_0]/\bar{n}_h$, where $E_0$=-0.467775 is the energy at zero doping,
taken from Ref.~\onlinecite{Sandvik1997}.
We compare the ground state hole energies at $\bar{n}_h$=1/8 obtained by various methods in Table I.
The hole energy we obtained is $E_{\rm hole}^{\rm fPEPS}$=-1.619, which is slightly lower than the hole energy obtained from DMRG calculation,\cite{Corboz2011}
$E^{\rm DMRG}_{\rm hole}$=-1.612 with $\chi$ extrapolated to $\infty$.
The result is also lower than the one from iPEPS SU calculations $E^{\rm iPEPS}_{\rm hole}$=-1.593, \cite{Corboz2011} which was obtained by extrapolating bond dimension $D$ to $\infty$.
The recent iPEPS full update calculations with $D$=14
give the hole energy $E^{\rm iPEPS}_{\rm hole}$=-1.578 for $\bar{n}_h$=0.12,\cite{Corboz2014}
which is expected to has lower (more negative) hole energy than that of $\bar{n}_h$=1/8.
As a comparison, the recent variational QMC simulation gives $E^{\rm QMC}_{\rm hole}$=-1.546 \cite{Hu2012}. We note that the ground state energy obtained in this work is significantly lower than the ground state energies obtained by DMRG and iPEPS calculation before extrapolation, which are close to $E^{\rm QMC}_{\rm hole}$.

\begin{table} [btp]
\caption{Compare the ground state hole energies of t-J model obtained by different
methods at hole doping $\bar{n}_h$=1/8 (except for the full update iPEPS calculation where $\bar{n}_h$=0.120).
The parameters $t$=1, $J/t$=0.4 are used in all calculations.
}
\begin{tabular}{  c c c c}
\hline\hline
method & parameter & $\bar{n}_h$ & Hole Energy  \\
\hline
VQMC + Lanczos  \cite{Hu2012}    & $p$=2              & 1/8 & -1.546\\
iPEPS simple update\cite{Corboz2011}  & $D$ $\to \infty$    & 1/8 & -1.593\\
iPEPS full update \cite{Corboz2014}   & $D$=14             & 0.120  & -1.578 \\
DMRG   \cite{Corboz2011}              & $\chi\to\infty$  & 1/8 & -1.612 \\
This work           & $D$=12             & 1/8 & -1.619\\
\hline
\hline
\end{tabular}
\label{tab:t-J}
 \end{table}

We now take a closer look at the t-J model.
The hole density and magnetization of the 4$\times$4 lattice obtained from fPEPS are compared with those obtained by diagonalization method\cite{Hellberg1999} in Fig.~S1 of the Supplementary materials (SM\cite{SM}). They are in remarkably good agreement.
The calculations suggest that ground state of the t-J model with hole filling of $\bar{n}_h=\frac{1}{8}$
on the 4$\times$4 lattice is in an uniform phase with virtually no local magnetization (the local magnetization is less than 10$^{-4}$),
which is in good agreement with the conclusions of Ref.\onlinecite{Hellberg1999}.

Would the uniform state be stable when the size of the system is increased? Unfortunately, the hole distribution become non-uniform when increase the size of the system. For the 4$\times$8 system, the ground state of the system is not
uniform any more. The holes clusters are more localized in the center of the lattice without local magnetic order [see Fig.~S2(a) in the SM\cite{SM}].
When the lattice size is further increased to 6$\times$8 and larger, the holes form stripes along the diagonal direction [see Fig.~S2(b,c) in the SM\cite{SM}].

\begin{figure} [tbp]
		\begin{center}
		\includegraphics[width=0.45\textwidth]{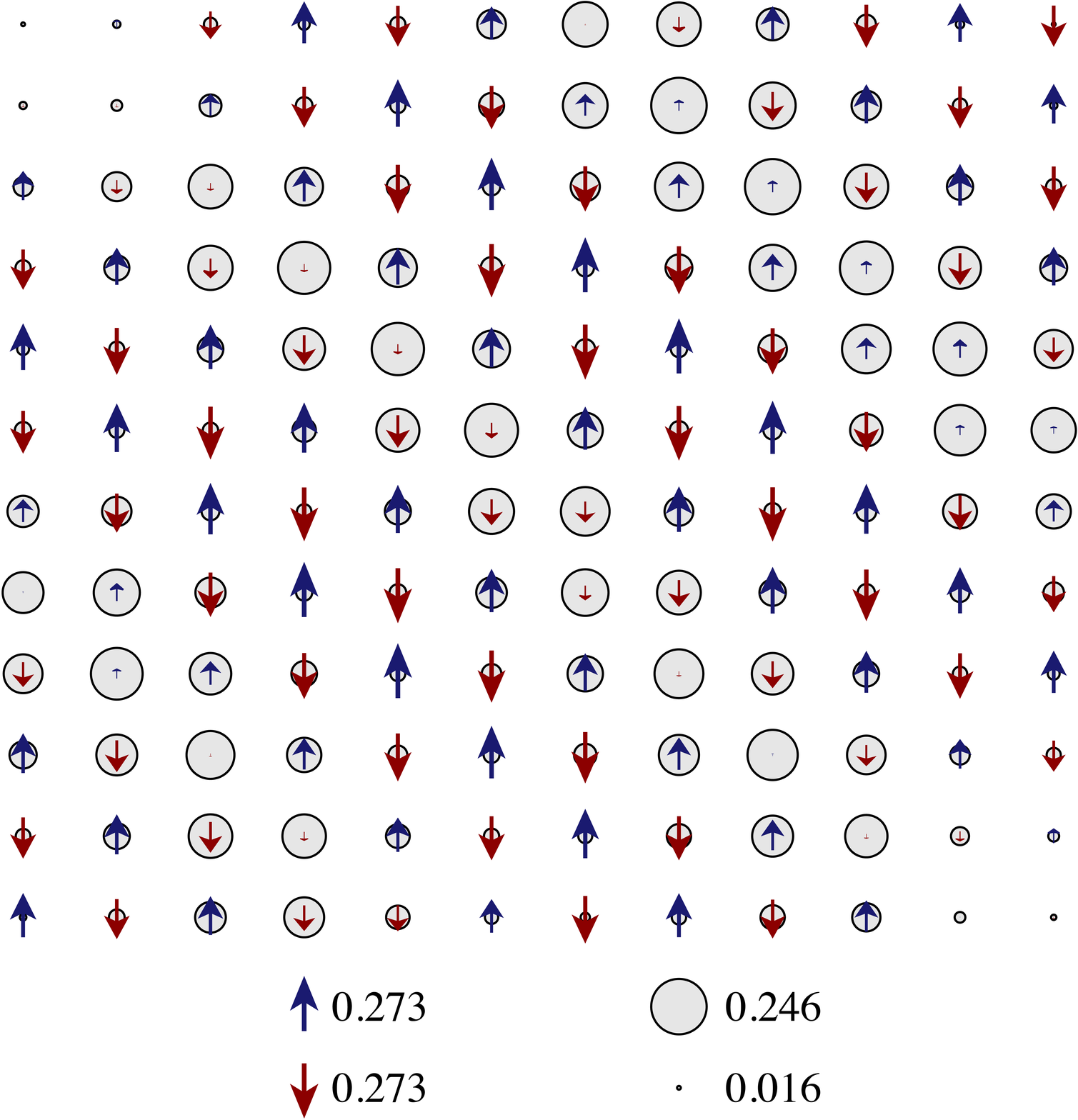}
		\caption{The ground state hole density and spin moment on the 12$\times$12 lattice.
The diameter of the circles represents the magnitude of holes density and length
of the arrow represents the local magnetic moments.
}
\label{pic:12_12}
		\end{center}
\end{figure}

Figure~\ref{pic:12_12} depicts the ground state hole distribution and local magnetization of the 12$\times$12 lattice.
The sizes of the circles and arrows represent the magnitude of the hole density and local magnetic moments. The systems show clear stripe order along diagonal direction on the antiferromagnetic background, with a $\pi$ phase-shifted magnetic order across the domain wall.\cite{White1998}

\begin{figure} [tbp]
		\begin{center}
		\includegraphics[width=0.48\textwidth]{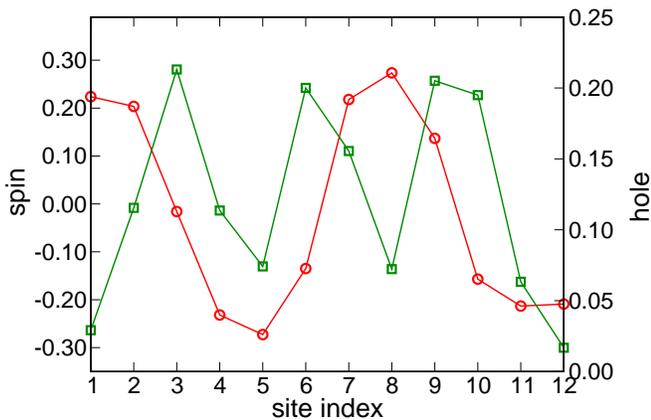}
		\caption{The average hole density $\langle n^{h} \rangle$ (green squares) and spin structure function (red circles)
along the diagonal direction on
the $12\times 12$ lattice.
}
\label{pic:spin_hole}
		\end{center}
\end{figure}

To investigate the structure of the diagonal stripe states, we plot the hole density $\langle n^{h}_{i,j} \rangle$=1-$\langle n_{i,j} \rangle $, where $\langle n_{i,j} \rangle$ is the average number of electrons on site $(i,j)$,
and staggered magnetization $(-1)^{i-1}S_{i,j}^z$ along the diagonal direction perpendicular to the stripes in Fig.~\ref{pic:spin_hole}.
The staggered magnetization (red line) shows a period of 8, whereas the hole density shows a period of 4. The site-centered nature of the hole stripes is evident from the hole density $\langle n^{h}_{i,j} \rangle$ (green line). These hole and spin pattern with period of 4 and 8, which are robust for different size of systems, can be used to explain why the stripe order cannot be stabled in the small 4$\times$ 4 and 4$\times$ 8 systems, which are too small to accommodate such stripes. The stripes has a hole filling $\rho_l$=$W\cdot \bar{n}_h$=0.5, where $W$ is the width of the domain wall, i.e., half filling.\cite{White1998}

To further test the robustness of the diagonal stripes against the size of the system, we simulated on lattices of different sizes (see Table.~S1 in the SM\cite{SM}), and aspect ratios. Except in some extreme cases, e.g. the width of the lattice is less than 4, we always obtain the diagonal stripes with similar spin and hole distribution structures.
These results clearly demonstrate the diagonal stripes ground states are robust against the size and shape of the lattices.

The t-J model at $\bar{n}_h$=1/8 has been intensively investigated by various methods, and the results are still highly controversial.\cite{Han2001,Sherman2003,Capello2008} On the one hand,
the recent variational quantum Monte Carlo (QMC) simulations combined with few Lanczos steps \cite{Hu2012} suggest that the ground state at $\bar{n}_h$=1/8 is homogeneous without stripes order. Its hole energies are very close to those obtained from DMRG \cite{White1998,White2000PSAndStripe,White2004,White2009ttJ} and recent iPEPS calculations\cite{Corboz2014}. On the other hand, DMRG calculations show that the ground state has stable stripes. \cite{White1998,White2000PSAndStripe,White2004,White2009ttJ,Corboz2014}

Our results support that the stripe phase is stable, which is in agreement with the DMRG calculations.
\cite{White1998,White2000PSAndStripe,White2004,White2009ttJ,Corboz2014}
In DMRG calculations, the stripes are further characterized as the site-centered vertical stripes,
and the width of the stripes are 4 at $\bar{n}_h\sim 1/8$,\cite{White1998}
which are also in good agreement with our results.
However, in our calculations, the stripes are along the diagonal direction in contrast to the vertical stripes obtained from DMRG calculations. This discrepancy may come from different BCs used in the calculations. In the DMRG calculations, a periodic BC in $y$-axis, and an open BC in $x$-axis are used, which favors the vertical stripes along the $y$ direction.\cite{White1998} One possible way to clarify this problem would be to perform DMRG calculations with periodic BC along the diagonal direction, to enforce a ground state with diagonal stripes, and compare the energy with that of vertical stripes.

Very recently, the iPEPS calculations~\cite{Corboz2011,Corboz2014} 
also suggest that the ground state is a vertical stripe phase, where the width of stripes and stripe hole filling depend on the exchange parameter $J$. At $J/t$=0.4, and $\bar{n}_h\sim$1/8, they obtain stripe filling $\rho_l \sim$ 0.5, which is in agreement with DMRG calculations. They also compare the energies of diagonal stripes and vertical stripes, and it has been found that the diagonal stripes have somewhat higher energies than the vertical stripes.~\cite{Corboz2011,Corboz2014}
However, the diagonal stripes obtained by the iPEPS calculations are very different from our cases. In Ref.\onlinecite{Corboz2014}, the $L \times L$  (for $L$ up to 11) supercells were used to obtain the diagonal stripe phase, but only $L$ (instead of $L^2$) tensors were independent.
With these constrains, the resulting diagonal stripes are insulating with filling $\rho_l$=1 holes per unit length, compared to $\rho_l$=0.5 holes per unit length in our calculations.
We remark that our calculations are unbiased in the sense that we do not have any constrains on the tensors.
All $L_1 \times L_2$ tensors are independent and free to change during the optimization.
We always obtain the same stripe ground states for randomly chosen initial states.
Another important difference between our method and the iPEPS method is that the iPEPS method directly works in the thermodynamic limit, which require extremely large $D$ to converge the results. In practice, such large $D$ is infeasible, and therefore, the final results rely heavily on the extrapolation on the bond dimension $D$. In fact, it has been found that the energy differences between the uniform state and the stripes states become smaller and smaller with increasing $D$,\cite{Corboz2014} Therefore, no definite conclusion can be made based on their current numerical results.
On the contrary, we work on finite systems, where the results can be fully converged with the given $D$. We then extrapolate the results to the thermodynamic limit by the well established finite-scaling method \cite{Finite-Size_Scaling}.

As a comparison, we also calculate the anisotropic t-J model with ${t_x}/{t_y}$=0.85, and ${J_x}/{J_y}$=$0.85^2$ following Ref.\onlinecite{Corboz2014}. We show the result of $J_x=0.4t_x$, $t_y=0.85t_x$ and $J_y=0.289t_x$ for different sizes in Fig.~S3 of the SM\cite{SM}. Compared with the isotropic case, the stripe orients along the bonds with stronger couplings, which is in agreement with previous results,~\cite{Corboz2014,Kampf2001,Chou2010}. The results can be understood as the kinetic energy can be effectively lowered
by hopping along this directions.
The anisotropic interaction converts the site-centered diagonal stripes into bond-centered vertical ones in these simulations.

\begin{figure} [tbp]
		\begin{center}
          \includegraphics[width=0.48\textwidth]{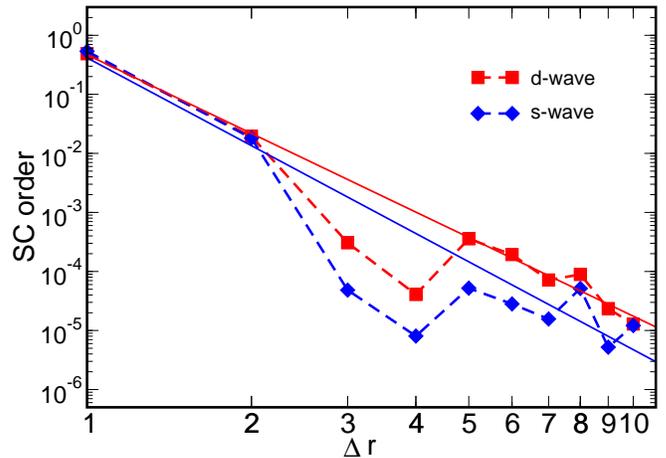}
       		\caption{The pair correlation function for the d-wave and the s-wave superconductivity on $12\times 12$ lattice. We fix  $\boldsymbol { r } _ { i }=(6,2)$ and scan $\boldsymbol { r } _ { j }=(6,2)$ to $\boldsymbol { r } _ { j }=(6,12)$, where $r$ is defined as $|\boldsymbol { r } _ { j } -\boldsymbol { r } _ { i }|$.
       } \label{pic:SCorder}
		\end{center}
\end{figure}

To further investigate the relationship between the stripe order and the superconductivity, we calculate the hole
pair correlation functions, which are defined as,
\begin{equation}\label{d-pairing}
P _ { \mathrm {s, d } } ( i , j ) = \left\langle \Delta _ { \mathrm {s, d } } ^ { \dagger } \left( \boldsymbol { r } _ { i } \right) \Delta _ { \mathrm {s, d } } \left( \boldsymbol { r } _ { j } \right) + \Delta _ { \mathrm {s, d } } \left( \boldsymbol { r } _ { i } \right) \Delta _ { \mathrm {s, d } } ^ { \dagger } \left( \boldsymbol { r } _ { j } \right) \right\rangle
\end{equation}
where s, d denote the s- or d-wave paring.
The superconductivity order parameter $\Delta_{\rm s, d}(\boldsymbol { r } _ { i } )$ is defined as,
\begin{equation}
\begin{aligned} \Delta _ { \mathrm {s, d } } \left( \boldsymbol { r } _ { i } \right) = & \sum _ { \pm } \frac { 1 } { 2 } \left\{ \left( c _ { i \uparrow } c _ { i \pm \hat { x } \downarrow } - c _ { i \downarrow } c _ { i \pm \hat { x } \uparrow } \right) \pm \left( c _ { i \uparrow } c _ { i \pm \hat { y } \downarrow } \right. \right. \\ & - c _ { i \downarrow } c _ { i \pm \hat { y } \uparrow } ) \} \end{aligned}
\end{equation}
with $\boldsymbol { r } _ { i }$ being the coordinate at site i, with ``$+$'' for s- wave and ``$-$'' for d-wave paring.
In Fig.~\ref{pic:SCorder}, we show both the s- and the d-wave pair correlation functions $P _ { \mathrm {s,d } } ( i , j )$ with $\boldsymbol { r } _ { i }$ fixed at (6,2) and $\boldsymbol { r } _ { j }$ changed from (6,2) to (6,12) in the 12$\times$ 12 lattice. To obtain highly accurate results, $D_c$=6$D$ is used to calculate the correlation functions.
The dips in the correlation functions around $r \sim$ 3 - 5 are presumably related to the hole stripe structures.
As shown in the figure, the superconductivity order for both s-wave and d-wave pairing
in the t-J model at $\bar{n}_h$=1/8 decay rather quickly with distance.
Even though the pair correlations can be fitted roughly by power law decay functions, i.e.,
$P_{\mathrm{s,d}}(r) \sim r^{-\alpha}$, with $\alpha \sim$ 4.9 and 4.4 for s-wave and d-wave pairing respectively,
since $\alpha \gg$1, the long range order of superconductivity is suppressed at  $\bar{n}_h$=1/8.


To summarize, we investigate the ground state of t-J model at hole doping $\bar{n}_h=1/8$, using the recently developed
highly accurate fPEPS method. We obtain the most competitive ground state hole energy. We find that the ground state has stable stripes along the diagonal direction, with stripe hole filling $\rho_l$=0.5. These results partially agree with recent DMRG and iPEPS calculations, in the sense that the ground state has stable stripes, except that in above calculations the stripes are vertical. We further show that the long range order of superconductivity is suppressed in this phase.
The work provides a new scenario of the ground state of the long standing open problem.

This work was funded by the National Key Research and Development Program of China (Grant No. 2016YFB0201202), the Chinese National Science Foundation (Grants No. 11774327, No. 11874343, No. 11474267), and the Strategic Priority Research Program (B) of the Chinese Academy of Sciences (Grant No. XDB01030200). China Postdoctoral Science Foundation funded project (Grant No. 2018M632529). It is also partially funded by the U.S. Department of Energy, Office of Science, Office of Advanced Scientific Computing Research, Scientific Discovery through Advanced Computing (SciDAC) program.(C.Y.) The numerical calculations have been done on the USTC HPC facilities and
the National Energy Research Scientific Computing (NERSC) center.


\end{document}